\documentclass[twocolumn,showpacs,amsmath,aps,pra,amssymb,nofootinbib,superscriptaddress]{revtex4-1} 
\usepackage[T1]{fontenc}
\usepackage[latin9]{inputenc}
\usepackage{times}
\usepackage{color} 
\usepackage{amssymb,amsmath}
\usepackage{amsbsy}
\usepackage{graphicx}
\usepackage{bm}
\usepackage{epstopdf}
\usepackage{float}
\usepackage{adjustbox}

\usepackage[unicode,breaklinks]{hyperref}
\hypersetup{
    unicode=true,
    a4paper=true,
    plainpages=false, 
    colorlinks=true,
    linkcolor=blue,
    citecolor=blue,
    filecolor=black,
    urlcolor=blue
}
\begin{document} 
\title{Dynamics of a quenched spin-1 antiferromagnetic condensate in a harmonic trap}
\author{L.~M.~Symes}
\author{D.~Baillie}
\author{P.~B.~Blakie}  
\affiliation{The Dodd-Walls Centre for Photonic and Quantum Technologies, New Zealand}
\affiliation{Department of Physics, University of Otago, New Zealand}

\begin{abstract}
 In this paper we consider a recent experiment [Kang \textit{et al.}~PRA {\bf 95} 053638 (2017)] in which an antiferromagnetic spin-1 condensate of $^{23}$Na atoms was quenched from the easy-axis polar phase  into the easy-plane polar phase. We simulate the phase transition dynamics using a classical field method with noisy initial conditions and find good agreement to the experimental observations. We analyse the late-time dynamics in terms of phase ordering using the nematic order parameter that we have developed in earlier work  [Symes \textit{et al.}~PRA {\bf 96}, 013602 (2017)]. We show that these experiments  are able to explore the early stages of universal phase ordering where the domains grow diffusively.  
\end{abstract} 
  
\maketitle
\section{Introduction}
Spin-1 condensates \cite{Stenger1998a,Ho1998a,Ohmi1998a} with antiferromagnetic interactions prefer to order into spin-nematic phases \cite{Kawaguchi2012R}. Such phases have a vanishing average spin-density $\langle f_a\rangle=0$, and can be characterized with the  nematic tensor 
\begin{align}
\mathcal{N}_{ab}=\tfrac{1}{2}\langle f_af_b+f_bf_a\rangle,\label{nematicN}
\end{align}
 where $f_{a\in\{x,y,z\}}$ are the spin matrices for spin-1.  The ground states of the antiferromagnetic system have an axially symmetric nematic tensor (uniaxial nematic) with a preferred axis characterized by a director $\vec{u}$ in spin-space \cite{Zibold2016a}.  In the presence of a quadratic Zeeman shift of the sublevels, $E_Z=q\langle f_z^2\rangle    =q(1-u_z^2)$, where $q$ is the quadratic Zeeman energy, the system experiences spin anisotropy. For $q>0$ the easy-axis polar (EAP) phase occurs where $\vec{u}\parallel \hat{z}$, while for $q<0$ the easy-plane polar (EPP) phase occurs where $\vec{u}\perp\hat{z}$.

By suddenly changing $q$, an antiferromagnetic spinor condensate can be taken from the EAP to the EPP.  Here the system undergoes a symmetry-breaking phase transition since the EPP breaks the in-plane rotational symmetry.  Following the quench, causally disconnected spatial domains of order develop, where in each domain the symmetry is broken independently with a randomly selected phase. In previous work we developed a theoretical description of this phase transition in which we defined an order parameter to quantify the spin-nematic degrees of freedom \cite{Symes2017b}. In that work we studied the late-time phase transition dynamics in a uniform two-dimensional (2D) system, demonstrating that the phase ordering is universal: the characteristic size of the domains develop with the diffusive growth law $L(t)\sim [t/\ln(t)]^{1/2}$ and  dynamic scaling holds (i.e.~correlation functions of the order parameter are time-independent when lengths are scaled by $L$) \cite{Bray1994}. The observation of universal phase-ordering dynamics is of significant interest in spinor condensates (e.g.~see \cite{Sadler2006a,Mukerjee2007a,Guzman2011a,Kudo2013a,Witkowska2013a,De2014a,Hofmann2014,Kudo2015a,Williamson2016a,Williamson2017a}) where a rich ground state phase diagram exists and is conveniently explored by varying parameters such as the quadratic Zeeman energy. Experimental studies of the EAP-to-EPP phase transition dynamics have been performed in quasi-one-dimensional \cite{Bookjans2011b,Vinit2013a} and quasi-2D \cite{Kang2017a} systems. The quasi-2D work was performed in a large ($\sim8\times10^6$  atoms) condensate and presents results for long evolution times  $\sim10^2\,t_s$ following the quench, where $t_s$ is the time scale associated with spin dynamics. The experimental analysis focused on the evolution of the axial spin density. This is not associated with the order of the system, but dynamic instabilities arising from the quench lead to the production of transverse magnetization and subsequently axial magnetic fluctuations. The experimental analysis showed that the magnetic fluctuations evolved from large to small length scales consistent with a direct turbulent cascade.   

Here we simulate these recent experiments using a classical field technique. This provides an important test of this method for application to nonequilibrium dynamics of spinor condensates. This also allows us to view the system dynamics in terms of phase ordering that is complementary to the turbulence analysis presented in Ref.~\cite{Kang2017a}. To do this we compute the correlation functions of the spin and superfluid order parameters, quantifying how these grow following the quench. Importantly we show that in these experiments the system size and the time scales of observation were sufficient to observe the universal diffusive growth of the ordered domains. Finally, while the EPP order is difficult to measure with current experimental tools, the defects of the order parameters are half-quantum vortices (HQVs) that have been observed \cite{Seo2015a,Seo2016a,Kang2017a}. Our results show that the scaling of the number of defects can also be used to reveal the growth of order in the system. We hope that these results will motivate experimental studies to quantify the late-time phase transition dynamics of spinor condensates in quasi-2D traps.

We briefly outline the paper. In Sec.~\ref{Sec:System} we introduce our system, including the formalism and techniques we use to simulate the quench, and review the relevant order parameters for the EPP phase. We present our results in Sec.~\ref{Sec:Results}. We initially focus on comparing to several quantities measured in experiments, such as the initial state decay, and how axial spin develops. An analytic model based on Bogoliubov theory is developed for the initial decay and compared to our results. We then consider how the EPP order emerges locally and then how it extends across the system. To do this we characterize the domains using correlation functions of the spin and superfluid order parameters, and use these to quantify the domain growth following the quench. We also compute the number of HQVs in the system and compare this to the length scales obtained from the correlation functions. Finally we conclude in Sec.~\ref{Sec:Conclusion}.
  
\section{System and Formalism}\label{Sec:System}
The evolution of a spin-1 condensate is given by the Gross-Pitaevskii equation (GPE)
\begin{equation}
i\hbar\frac{\partial\bm{\psi}}{\partial t}=\mathcal{L}_{\mathrm{GP}}\bm{\psi},\label{GPE}
\end{equation}
where
\begin{equation}
\mathcal{L}_{\mathrm{GP}}\equiv-\frac{\hbar^2\nabla^2}{2M}+V_{\mathrm{trap}}(\mathbf{x})+qf_z^2+g_nn+g_s\bm{F}\cdot\bm{f},
\end{equation}
and $\bm{\psi}=(\psi_1,\psi_0,\psi_{-1})^T$, with $\psi_m$ the field in sublevel $m$.
The first nonlinear term describes the density dependent interactions, with coupling constant $g_n=4\pi a_n\hbar^2/M$, where  $n\equiv\bm{\psi}^\dagger\bm{\psi}$ is the total density.  The second nonlinear term describes the  spin-dependent interactions,  with coupling constant $g_s=4\pi a_s\hbar^2/M$,  where  ${F}_a\equiv \bm{\psi}^\dagger{f}_a\bm{\psi}$  is the $a$-component  ($a=\{x,y,z\})$ of spin density. For $^{23}$Na we take, except where otherwise stated, the scattering lengths to be $a_n=51.1\,a_0$ and $a_s=0.823\,a_0$, where $a_0$ is the Bohr radius \cite{Black2007a}.

\subsection{Initial state}\label{SecIS}
\begin{figure}[t] 
\centering
\includegraphics[trim=5 0 0 0,clip=true,width=1.02\linewidth]{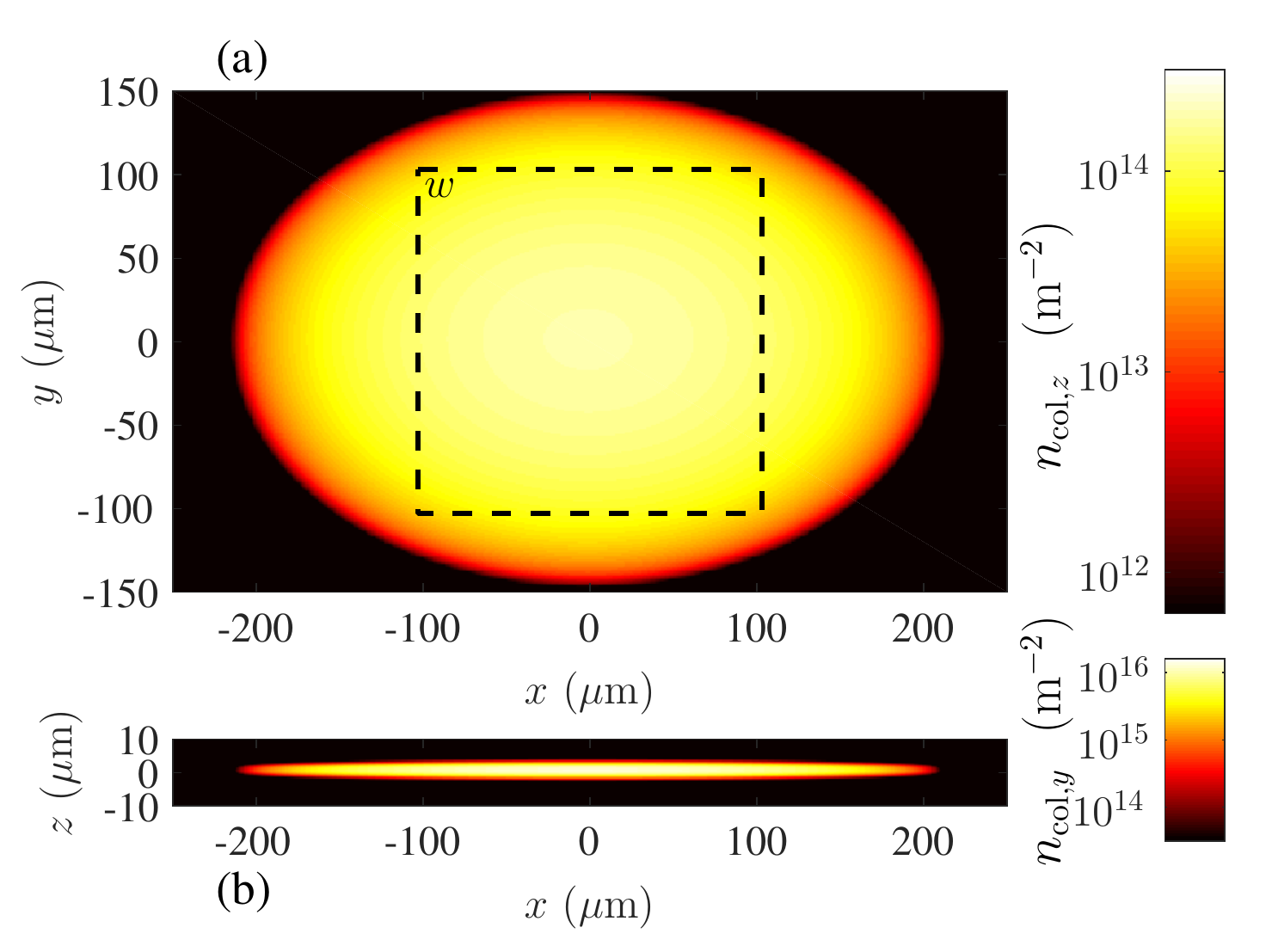}
\caption{Initial condensate density profile. (a) $z$-column density  $n_{\mathrm{col},z}\!\equiv\!\int dz\,|\psi_g(\mathbf{x})|^2$ and (b) $y$-column density  $n_{\mathrm{col},y}\!=\!\int dy\,|\psi_g(\mathbf{x})|^2$ of the GPE ground state. The black square in (a) marked with $w$ indicates the boundary of the  $206\,\mu$m$\,\times\,206\,\mu$m region used in experiments and in this work to analyze spatial correlations.
 }
\label{fig:GS}
\end{figure}

We perform simulations corresponding to the experiment and take a condensate with $N_c=8\times10^6$   atoms prepared in an oblate harmonic trap 
$V_{\mathrm{trap}}(\mathbf{x})=\frac{1}{2}M\sum_{a}\omega_a^2x_a^2$, 
with frequencies $(\omega_x, \omega_y, \omega_z)/2\pi =  (3.8, 5.5, 400) \; \mathrm{Hz}$ and $q>0$.  The initial condensate state in this regime is in an EAP state and is found as the ground state solution of the time-independent GPE $\mu\bm{\psi}_{\mathrm{EAP}}=\mathcal{L}_{\mathrm{GP}}\bm{\psi}_{\mathrm{EAP}}$, where $\mu$ is the chemical potential. The EAP state only has occupation of the $m_F=0$ sublevel, i.e.
\begin{align}
\bm{\psi}_{\mathrm{EAP}}(\mathbf{x})=\begin{pmatrix}0 \\ \psi_g(\mathbf{x}) \\0\end{pmatrix}.
\end{align}
 We find the $\psi_g$ orbital as the solution of the scalar GPE with scattering length $a_n$ using the Newton-Krylov method. Our solution (see Fig.~\ref{fig:GS}) has chemical potential $\mu/h = 939\,$Hz, and peak spin energy  $g_sn_{\mathrm{peak}}/h= 14.3\,$ Hz, where  $n_{\mathrm{peak}}=|\psi_g(\mathbf{0})|^2$ is the peak density occurring at the trap center. The $1/e$-density widths of the condensate are $(R_x,R_y,R_z) =(173,120,1.75)\,\mu$m. The spin healing length at the trap center is $\xi_s = \hbar/\sqrt{2M g_s n_{\mathrm{peak}}} =3.9 \; \mu$m, which is larger than $R_z$, so that spin textures are frozen out in this direction and the system can be regarded as being quasi-2D with respect to the spin-degrees of freedom.  

To simulate the system dynamics it is essential to include noise in the initial state to seed dynamical instabilities that occur once $q$ is quenched.  
We account for the initial state noise using the truncated Wigner formalism \cite{cfieldRev2008}:
\begin{align}
\bm{\psi}(\mathbf{x},t=0)&=\bm{\psi}_{\mathrm{EAP}}(\mathbf{x}) +\bm{\delta}(\mathbf{x}),\label{3Dinitialstate}
\end{align} 
where   $\bm{\delta}(\mathbf{x})$ denotes a noise vector. Ideally, the noise vector would be constructed using the Bogoliubov quasiparticle modes, but this is a challenging calculation for a three-dimensional spinor system. Here we add noise using the single-particle basis such that the component fields of $\bm{\delta}$ are
\begin{align}
\delta_m(\mathbf{x})=\sum^\prime_j\alpha_{m,j}\,\phi_j(\mathbf{x}),\label{deltanoise}
\end{align}
where $\{\phi_j(\mathbf{x})\}$ are the single-particle harmonic oscillator basis modes for the trapping potential with respective energy eigenvalues $\{\varepsilon_j\}$, and the prime on the summation indicates that it is restricted to basis states with $\varepsilon_j<k_BT$ where $T$ is the system temperature. 
The $\{\alpha_{m,j}\}$ are independent complex Gaussian random numbers, with $\langle\alpha_{m,j}\rangle=0$ and variances given by
\begin{align}
\langle|\alpha_{m,j}|^2\rangle=\left\{\begin{array}{cl} \frac{1}{2} \qquad & m=\pm1, \\ \bar{n}_j+\frac{1}{2} \qquad & m=0,\\ \end{array}\right.\label{alpha_noise}
\end{align}
where $\bar{n}_j=(e^{\epsilon_j/k_BT}-1)^{-1}$ is the mean thermal occupation.\footnote{Here $\langle\cdot\rangle$ indicates an average over stochastic realizations of the random variable $\alpha_{m,j}$.} The  $\tfrac{1}{2}$'s appearing in  (\ref{alpha_noise}) represent the vacuum noise for the symmetrically ordered Wigner representation.
 The initial state \eqref{3Dinitialstate} is thus a condensate with thermal excitations in the $m=0$ component and only vacuum noise in the $m=\pm1$ components. This corresponds to the initial condition prepared in experiments where the initial $q$ value is large and positive.
 The experiment reported that the initial thermal fraction was less than 10\%. Here we take $T=82\,\mathrm{nK}$ for the system temperature which is  $0.45T_c^0$, with $T_c^0$ being the ideal condensate temperature in the harmonic trap. At this temperature the noncondensate fraction for the ideal system is 9\%. Because we restrict the summation in Eq.~\eqref{deltanoise} to energies below $k_BT$, the average number of thermal particles included is approximately  $2\%$ of $N_c$. 
We have explored changing the cutoff and temperature in our simulations and find that it makes little difference to the subsequent dynamics.  This is because the initial dynamics following the quench are driven by the vacuum seeding in the $m=\pm1$ modes, and are not sensitive to the thermally occupied modes. The most important unstable modes have a typical energy scale ($\sim g_sn$) that is much lower than the energy cutoff used to construct the noise.

\subsection{Quench simulation}
To simulate the system dynamics we evolve the initial condition (\ref{3Dinitialstate}) according to the GPE (\ref{GPE}). We do this on a three-dimensional numerical grid using the S2 symplectic integrator introduced in Ref.~\cite{Symes2016a}. In each direction the spatial extent of the grid is $L_a$, i.e.~$-\tfrac{1}{2}L_a\!\le \!x_a\!<\!\tfrac{1}{2}L_a$, and is spanned by $N_a$ equally spaced points. For the results presented here we use
$({L}_x, {L}_y, {L}_z) =  (700 ,482 ,14.7)\,\mu$m  and  $(N_x, N_y, N_z) = (672, 464, 32)$. This choice ensures that the modes up to energy $k_BT$ are well represented on the grid. Simulation results are generally averaged over a number of independent trajectories, i.e.~individual solutions of the time-dependent GPE (\ref{GPE}), that differ by the sampling of initial noise.

\subsection{Nematic and superfluid order}\label{Sec:nematicorder}
 We briefly review the relevant order parameters for the EAP-to-EPP transition that we introduced in earlier work \cite{Symes2017b}. The key quantity for the EPP spin order is the in-plane nematic tensor
 \begin{align}
 Q\equiv \mathcal{N}_{2\times2}-\tfrac{1}{2}\mathrm{Tr}\{\mathcal{N}_{2\times2}\}I_2=\left(\begin{array}{cc} Q_{xx} & Q_{xy}\\ Q_{xy} & -Q_{xx}\end{array}\right), 
 \end{align}
 where $ \mathcal{N}_{2\times2}$ is the $xy$ submatrix of the full nematic tensor  (\ref{nematicN}), and $I_2$ is the ${2\times2}$-identity matrix. As defined above the in-plane nematic tensor is traceless and symmetric, with $Q_{xx}=\mathrm{Re}\{\psi_1^*\psi_{-1}\}$ and $Q_{xy}=\mathrm{Im}\{\psi_1^*\psi_{-1}\}$. It has the important property that $\mathrm{Tr}\{Q^2\}=0$ when the spin fluctuations are isotropic in the $xy$ plane, i.e.~in the  EAP phase. In the EPP phase the isotropy is broken and $\mathrm{Tr}\{Q^2\}>0$, thus this quantity acts as a spin-nematic order parameter for the phase transition.
  
The EPP state is of the general form
 \begin{align}
 \bm{\psi}_{\mathrm{EPP}}=\sqrt{\frac{n}{2}}e^{i\theta}\left(\begin{array}{c}-e^{-i\phi}\\ 0\\ e^{i\phi}\end{array}\right),\label{psiEPP}
 \end{align}
where the angle $\phi$ is associated with spin-nematic order (i.e.~the nematic director is $\vec{u}\sim\cos\phi\hat{\mathbf{x}}+\sin\phi\hat{\mathbf{y}}\,$) and $\theta$ is the global phase associated with superfluid order. Noting that $Q_{xx}\sim\cos2\phi$ and $Q_{xy}\sim\sin2\phi$, we see that $Q$ is insensitive to superfluid order. Thus to quantify superfluid order we introduce the in-plane component of the spin-singlet amplitude\footnote{The full spin-singlet amplitude for the spin-1 system is $\alpha=\psi_0^2-2\psi_1\psi_{1}$. Note that for the spin-1 system  $n^2=F^2+|\alpha|^2$, so the spin-singlet density $|\alpha|^2$ is complementary to the spin-density $F^2$ and is maximized for $g_s>0$.    }
\begin{align}
\alpha_\perp\equiv-2\psi_1\psi_{-1},\label{alphaperp}
\end{align}
which scales as $\alpha_\perp\sim e^{2i\theta}$.

To quantify the phase ordering dynamics following the quench, it is useful to introduce the correlation functions associated with these order parameters 
\begin{align}
G_\phi(\mathbf{x},\mathbf{x}')&\equiv\langle \mathrm{Tr}\{Q(\mathbf{x})Q(\mathbf{x}')\}\rangle,\label{Gphi}\\
G_\theta(\mathbf{x},\mathbf{x}')&\equiv\langle  \alpha^*_\perp(\mathbf{x})\alpha_\perp(\mathbf{x}')\rangle. \label{Gtheta}
\end{align}
These correlation functions reveal the spatial ordering of the spin-nematic order  with $G_\phi\sim\langle \cos2[\phi(\mathbf{x}')-\phi(\mathbf{x})]\rangle$, and the superfluid order with $G_\theta\sim\langle e^{i2[\theta(\mathbf{x}')-\theta(\mathbf{x})]}\rangle$ (see Ref.~\cite{Symes2017b} for more details).

\section{Results}\label{Sec:Results}

 \begin{figure}[t]
\centering 
\includegraphics[width=0.95\linewidth]{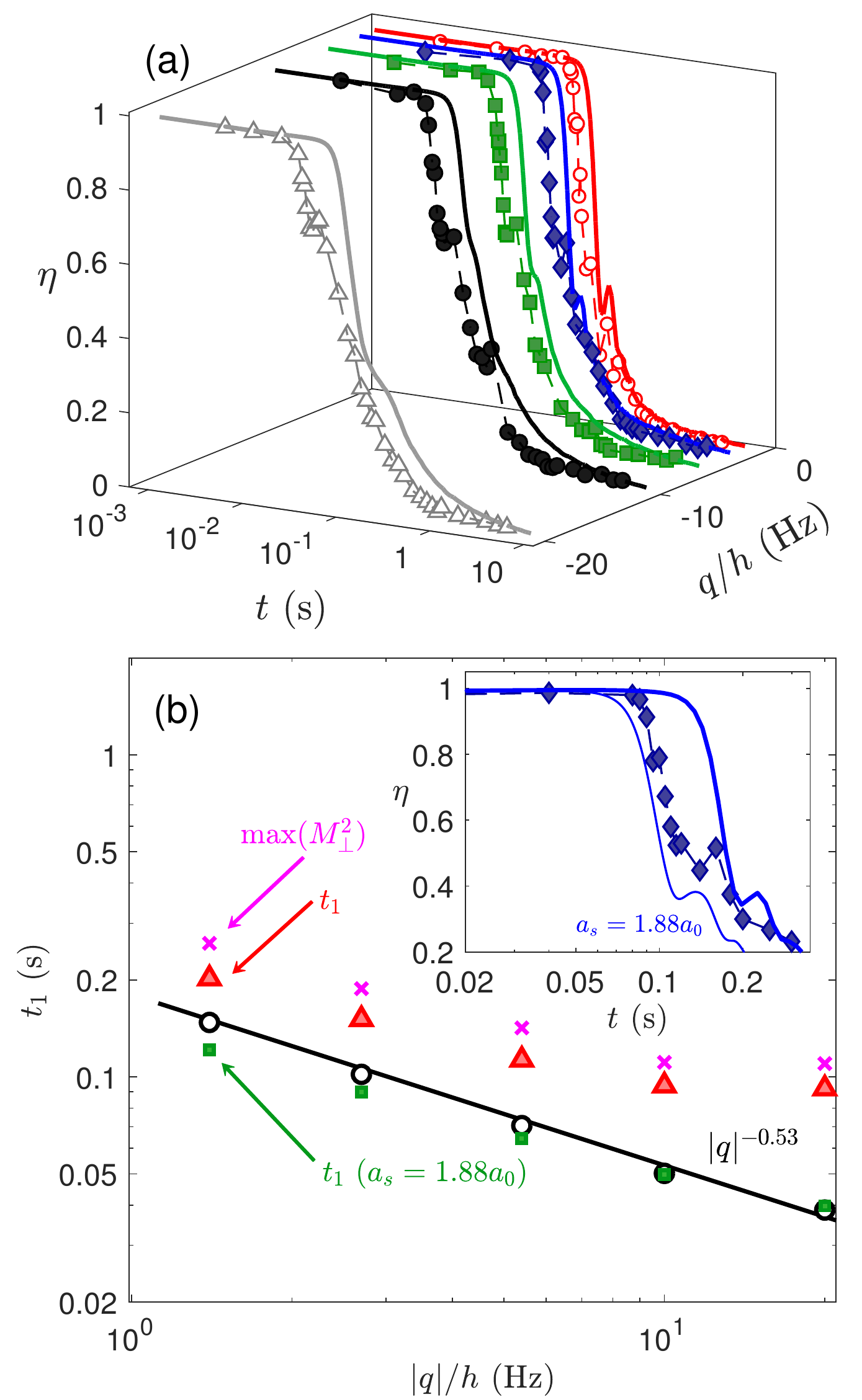} 
\caption{Decay of the initial condensate state as $q$ is varied. (a) Time evolution of $\eta$ for quenches to various $q$ values. Experimental results from Ref.~\cite{Kang2017a} are shown with symbols while our corresponding simulation results are given by solid lines. (b) The timescale $t_1$ at which $\eta$ reaches 0.8: experimental (black circles) and numerical results using $a_s=0.823\,a_0$ (red filled triangles) and $a_s=1.88\,a_0$  (green filled squares). The time at which $M_\perp^2$ is maximized in the numerical calculations is shown for reference (magenta crosses). The solid line shows a $|q|^{-0.53}$ fit to experimental data. Inset: Comparison of $\eta(t)$  from experimental results (diamonds) to the simulations using $a_s=0.823\,a_0$  from subplot (a) (thick blue line) and a single trajectory with $a_s=1.88\,a_0$  (thin blue line) for the case $q/h=-2.7\,$Hz. 
Results in (a) are  averaged over 4 trajectories, those in (b) use 64 trajectories, except the  $a_s=1.88\,a_0$ results which use 25 trajectories. The standard deviation in $t_1$ times between trajectories is comparable or smaller than the marker size. }
\label{fig:AFexpFig3}
\end{figure}

 \subsection{Decay of initial state} 
 We initiate the quench at $t=0$ by suddenly setting $q$ to a negative value.  This causes the initial EAP state to be unstable to transverse spin fluctuations that grow as spin exchange collisions transfer pairs of $m=0$ atoms into the $m=1$ and $-1$ sublevels. The depletion of the condensate following the quench can be characterized by the quantity 
\begin{align}
\eta(t)=N_0(t)/N_w(t),
\end{align}
where $N_0(t)=\int_w d\mathbf{x}\,|\psi_0|^2$, and $N_w(t)=\int_w d\mathbf{x}\,n$ are the $m=0$ and the total atomic population in the $w$-window, respectively [Note: the subscript $w$  denotes that the integration over the $xy$-plane is restricted to the square window of area $A_w=(206\,\mu$m$)^2$ indicated in Fig.~\ref{fig:GS}(a)].
Our results are compared to the experimental data in  Fig.~\ref{fig:AFexpFig3}(a), and are seen to have qualitatively similar behavior. The decay is also quantified by introducing a time $t_1$  defined as the time when the $m=0$ population decays to 80\% of its initial value, i.e.~$\eta(t_1)=0.8$. The experimental and simulation results for $t_1$ are compared in Fig.~\ref{fig:AFexpFig3}(b).  This reveals that  the simulation  $t_1$ times are longer than those measured in experiments, and follow different scaling with $q$ at higher $|q|$ values: the experimental results, even for the deepest quenches, scale as $t_1\sim|q|^{-0.5}$, while our simulations depart from this trend and saturate for quenches to $|q|/h>10\,$Hz. 

We have investigated adding more initial noise to our simulations through using a higher energy cutoff in (\ref{deltanoise}) and using higher temperatures. These changes were found to cause negligible differences to the $t_1$ time.  
 For the spin-dependent interaction we have used the value $a_s=0.823\,a_0$ reported by \cite{Black2007a} that was determined from spin oscillations of a sodium condensate with an uncertainty of 10\%.   
A more recent Feshbach study \cite{Knoop2011a} has characterised the scattering properties of sodium in detail. While this study did not directly measure the spin-dependent interaction, it can be determined from the scattering lengths they measured, yielding the larger value of $a_s=1.88\,a_0$.
In the inset to Fig.~\ref{fig:AFexpFig3}(b) we show an example trajectory calculated with $a_s=1.88\,a_0$, revealing a significantly reduced decay time. Results for $t_1$ averaged over trajectories are shown in Fig.~\ref{fig:AFexpFig3}(b). This comparison suggests that the spin-dependent scattering length is larger than the widely accepted value of $0.823\,a_0$. However, we cannot rule out that other factors we have not accounted for could play an important role, e.g.~heating in the experiment due to the microwave dressing used to control $q$, the effect of spatial gradients in the magnetic fields, or collective modes excited in the preparation of the condensate.  For the remaining results we present in this paper we use $a_s=0.823\,a_0$.

\begin{figure}[t] 
  \centering
  \includegraphics[trim=0 0 0 6,clip=true,width=0.825\linewidth]{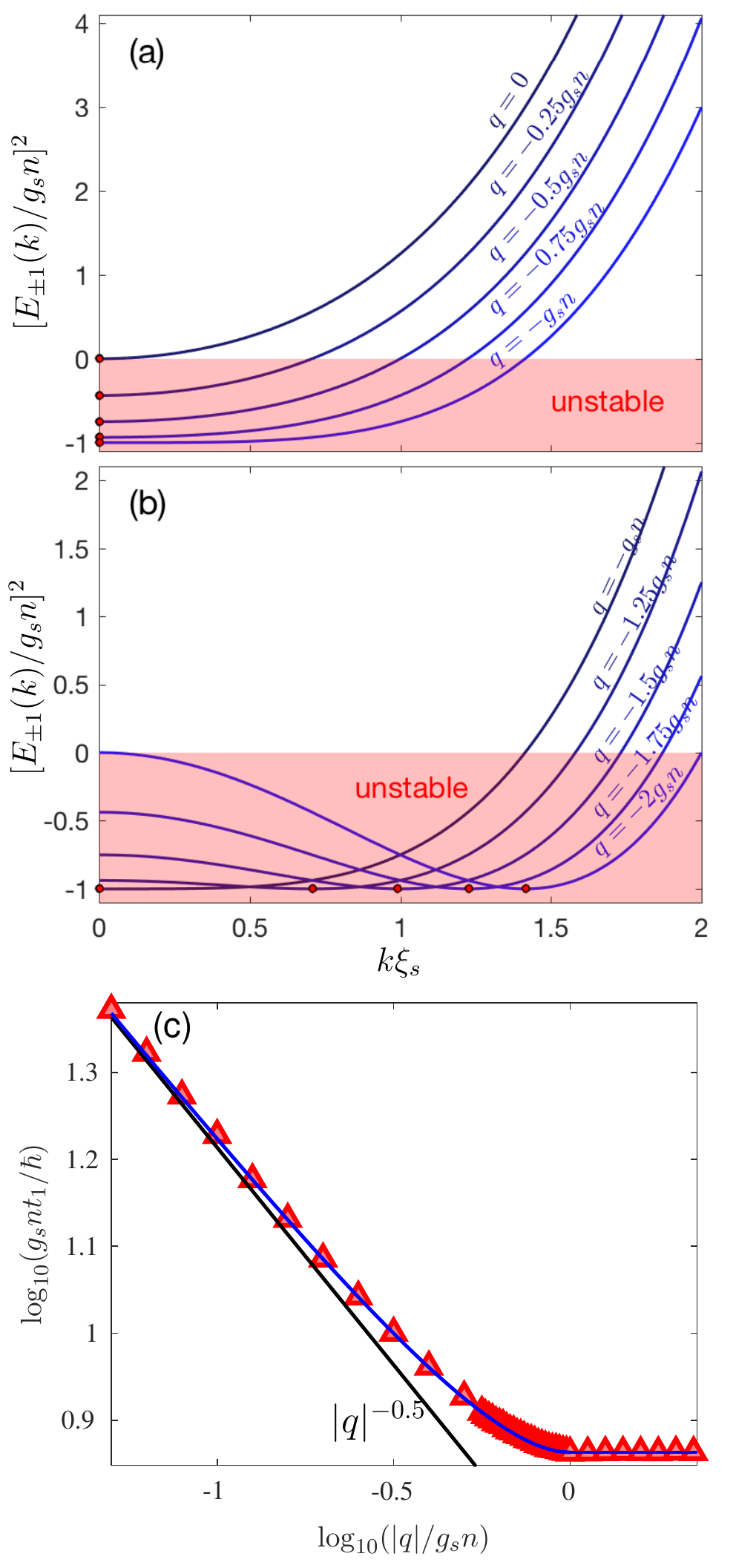}\vspace{-3mm}
\caption{\label{BogFig} Uniform Bogoliubov description of the EAP state instability at $q<0$. The square of the magnon dispersion (\ref{Emagnon}) is shown for (a) $ -g_sn\le q<0$ and (b)  $q\le-g_sn$. Where $[E_{\pm1}(k)]^2<0$   the modes are dynamically unstable. The most imaginary mode is indicated with a small red filled circle. For $ -g_sn< q<0$  this occurs at $k=0$ and $\mathrm{Im}\{E_{\mathrm{mi}}\}$ increases with $|q|$, whereas for $q<-g_sn$  it moves to non-zero $k$ but $\mathrm{Im}\{E_{\mathrm{mi}}\}$ is constant [see (\ref{EqEmi})]. (c)
Approximate Bogoliubov prediction for $t_1$ [i.e.~Eq.~\eqref{Uniformt1} with $c=7.3$ as a fit parameter] (blue curve) compared to the low $|q|$ estimate  $|q|^{-0.5}$ (black straight line), and the numerical results for   the trapped system (filled red triangles). Numerical results obtained by averaging over 100 trajectories. }\vspace{-6mm}
\end{figure}

To understand the initial decay dynamics we develop a simple model based on Bogoliubov theory for a uniform polar condensate of density $n$. The dispersion relation for the two branches of magnon excitations on top of the polar condensate are 
\begin{align} 
E_{\pm 1}(k) &= \sqrt{(\epsilon_k + q)(\epsilon_k + q + 2 g_sn)}.\label{Emagnon}
\end{align}
These branches are degenerate, and correspond to quasiparticles with momentum $\hbar k$ that have amplitude in both the  $m=1$ and $-1$ sublevels \cite{Kawaguchi2012R,Symes2014a}.
For $q<0$ this dispersion relation is imaginary for a range of $k$-values [see Figs.~\ref{BogFig}(a), (b)]. This signifies the onset of a dynamical instability where the population of these modes begins to grow exponentially.
The most unstable mode (i.e.~largest imaginary energy) has the energy
\begin{align}
E_{\mathrm{mi}}=\begin{cases}\sqrt{q(q+2g_sn)}& -g_sn<q<0 \\
ig_s n & q<-g_sn,\label{EqEmi}
\end{cases}
\end{align} 
and is indicated in Figs.~\ref{BogFig}(a), (b). 
Approximating the growth of population into the $m=\pm1$ sublevels by the most dynamically unstable mode, we have that the population of this mode will evolve as $\sim e^{2|E_{\mathrm{mi}}|t/\hbar}$, and thus we have $1-\eta(t)=\alpha e^{2|E_{\mathrm{mi}}|t/\hbar}$, where $\alpha$ is the growth rate prefactor. Solving for $\eta(t_1)=0.8$ yields the decay time as
\begin{align}
\log_{10}\frac{g_sn\,t_1}{\hbar}=\log_{10}c-\log_{10}\frac{|E_{\mathrm{mi}}|}{g_sn},\label{Uniformt1}
\end{align}
where $c=\frac12\ln (0.2/\alpha)$. 
This prediction for the decay time is shown in Fig.~\ref{BogFig}(c), notably predicting that the decay rate should saturate for $q\lesssim -g_sn$.

To apply the uniform Bogoliubov prediction to the trapped system, we need a value for $g_sn$ appearing in Eqs.~\eqref{Emagnon}-\eqref{Uniformt1}. We use $g_sn/h = 12.5\:\mathrm{Hz}$ as a fitting parameter, which is somewhat less than $g_sn_{\mathrm{peak}}/h = 14.3\:\mathrm{Hz}$ from the maximum three-dimensional density of the initial polar ground state shown in Fig.~\ref{fig:GS}, accounting for the inhomogeneous density distribution.
 
In Fig.~\ref{BogFig}(c) we compare results for our fully trapped simulations against result \eqref{Uniformt1} and find good agreement. Notably, we find that $t_1$ plateaus for $q<-g_sn$. For larger $|q|$ than shown in Fig.~\ref{BogFig}(c) we find that $t_1$ increases, due to beyond-Bogoliubov effects affecting the growth before $\eta(t) = 0.8$ is reached.

\subsection{Behavior of spin and spin-singlet densities}\label{Sec:spindensities}\vspace{-3mm}
The evolution of axial spin fluctuations following the quench was also measured. This corresponds to the observable\footnote{Since the total axial magnetization is conserved it remains at its initial value of zero, hence the expectation of $F^2_z$ in (\ref{dMz2}) corresponds to the fluctuations.}\vspace{-3mm}
\begin{align}
  \delta M_z^2 &\equiv \frac{A_w}{N_w^2}\int_wd\bm{\rho}\,\langle F^2_z(\bm{\rho})\rangle,\label{dMz2}
\end{align}
where $F_{a}(\bm{\rho})=\int dz\,F_{a}(\mathbf{x})$ is the column density of the $a$-component of spin density.

\begin{figure}[t] 
	\centering \includegraphics[width=0.9\linewidth]{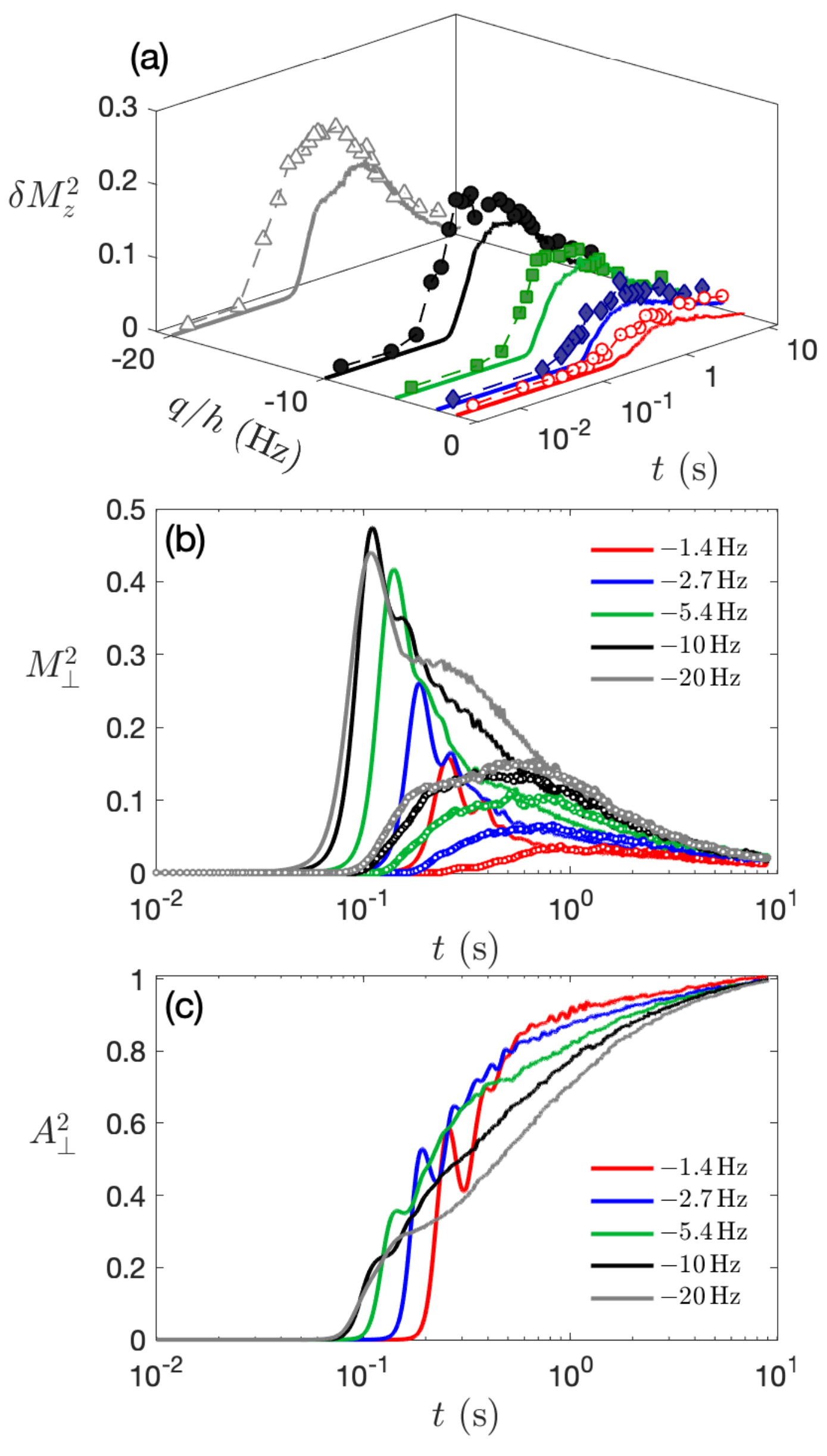}
    \caption{Local spin density behavior following the quench. (a) Axial spin density simulation results (solid lines) and  experimental results (symbols and dashed line). (b) Simulation results for the transverse magnetization (solid lines, increases as $q$ becomes more negative for $t\gtrsim 0.5\:\mathrm{s}$) with the axial simulation results from (a) shown for reference (lines with small circles). (c) Simulation results for the transverse spin- singlet density (decreases as $q$ becomes more negative for $t\gtrsim 0.5\:\mathrm{s}$).
}\label{fig:localspin} \end{figure}

A comparison of the experimental and simulation results is shown in Fig.~\ref{fig:localspin}(a). Similar to the results for $\eta$, the experimental results for $\delta M_z^2$ start growing earlier than is observed in the simulations. At long times ($t\to5\,$s) experiments observed $\delta M_z^2$  to approach a nonzero value independent of $q$, which was interpreted to be due to heating from the microwave field dressing and the evaporative cooling due to the finite trap depth. We do not account for these effects and our simulations according to Eq.~(\ref{GPE}) conserve total energy. However heating does occur in our simulations arising from the energy released by the quench\footnote{For the uniform system an energy per particle of $q$  is liberated by the quench, so for deeper quenches the system will thermalize to a higher $T$.} and we expect that at sufficiently late times (beyond the time scales considered here) $\delta M_z^2$ will approach a non-zero value that depends on $q$.

In addition we analyze the evolution of other densities not measured in the  experiment: 
\begin{align}
 M_\perp^2 &\equiv \frac{A_w}{N_w^2}\int_wd\bm{\rho}\,\langle  {F}_\perp^2(\bm{\rho})\rangle,\\
A_\perp^2 &\equiv \frac{A_w}{N_w^2}\int_wd\bm{\rho}\,\langle |\alpha_\perp(\bm{\rho})|^2\rangle,
 \end{align}
where  $\mathbf{F}_\perp=(F_x,F_y)$ denotes the transverse spin column density and $\alpha_\perp(\bm{\rho})$ is the column density of the in-plane spin-singlet amplitude. These quantities are relevant because the dynamically unstable modes directly generate transverse magnetization, while the local EPP order is revealed by the growth of $\alpha_\perp(\bm{\rho})$, noting that $\mathrm{Tr}\{Q^2\}=\tfrac{1}{2}|\alpha_\perp|^2$ (see Sec.~\ref{Sec:nematicorder}).

Our results for $ M_\perp^2$ are shown in Fig.~\ref{fig:localspin}(b), where they are compared to our results for $\delta M_z^2$. We observe that for each $q$ value $ M_\perp^2$  develops earlier than $ \delta M_z^2$. This is expected since transverse spin fluctuations are  directly generated from the unstable magnon modes. In general the transverse magnetization decays significantly before the  axial magnetization reaches its maximum. The transverse magnetization has a more prominent and well defined maximum. The time when this maximum occurs is similar to $t_1$ [see Fig.~\ref{fig:GS}(b)]. 

The results for $A_\perp^2$ are shown in Fig.~\ref{fig:localspin}(c). Unlike the spin density this quantity continues to grow as time increases  revealing the development of EPP order post-quench. For the quenches considered this quantity appears to saturate at late times to a value close to unity. We take this as a sign of the system approaching equilibrium.

\subsection{Growth of spatial order}
While the results for $A_\perp^2$ in Fig.~\ref{fig:localspin}(c) show that order is locally established in the system, it is of interest to understand how this order spatially extends across the system following the quench.   To quantify this we employ the order parameter correlation functions $G_\phi$ and $G_\theta$ defined in Eqs.~(\ref{Gphi}) and (\ref{Gtheta}). Here we adapt these to the trapped case by spatially averaging over the central window region [the box $w$ indicated in Fig.~\ref{fig:GS}(a)] and normalizing by the atom number in this region:
\begin{align}
G_\phi(\mathbf{r})&=\frac{2}{N_w^2}\int_w{d\bm{\rho}}{}\int_w{d\bm{\rho}'}{} \langle \mathrm{Tr}\{Q(\bm{\rho})Q(\bm{\rho}')\}\rangle\delta(\bm{\rho}-\bm{\rho}'-\mathbf{r}) ,\\
G_\theta(\mathbf{r})&=\frac{1}{N_w^2}\int_w{d\bm{\rho}}{}\int_w{d\bm{\rho}'}{}\langle \alpha^*_\perp(\bm{\rho})\alpha_\perp(\bm{\rho}')\rangle\delta(\bm{\rho}-\bm{\rho}'-\mathbf{r})  ,
\end{align}
with $Q(\bm{\rho})$ being $Q$ after integration along $z$.  These correlation functions are readily evaluated  using 2D Fourier transforms, which we denote as $\mathcal{F}$, e.g., $G_\phi(\mathbf{r})=\frac{2}{N_w^2}\mathcal{F}^{-1}\{\tilde{Q}_w(\mathbf{k})\tilde{Q}_w(-\mathbf{k})\}$, where $\tilde{Q}_w(\mathbf{k})=\mathcal{F}\{Q\}$ is the Fourier transform of $Q$ restricted to the window region. Similarly we can assess correlations in the spin order using
\begin{align}
    G_{F_\perp}(\mathbf{r})&=\frac{1}{N_w^2}\int_w{d\bm{\rho}}{}\int_w{d\bm{\rho}'}{} \langle \mathbf{F}_\perp(\bm{\rho})\cdot \mathbf{F}_\perp(\bm{\rho}')\rangle\delta(\bm{\rho}-\bm{\rho}'-\mathbf{r}) ,\\
G_{F_z}(\mathbf{r})&=\frac{1}{N_w^2}\int_w{d\bm{\rho}}{}\int_w{d\bm{\rho}'}{}\langle F_z(\bm{\rho})F_z(\bm{\rho}')\rangle\delta(\bm{\rho}-\bm{\rho}'-\mathbf{r}).
\end{align}
We note that $G_\theta(0)=G_\phi(0)=A_\perp^2$, $G_{F_\perp}(0)=M^2_\perp$ and $G_{F_z}(0)=\delta M^2_z$
as defined in Sec.~\ref{Sec:spindensities}.

\begin{figure}[H] 
\centering
\includegraphics[width=0.9\linewidth]{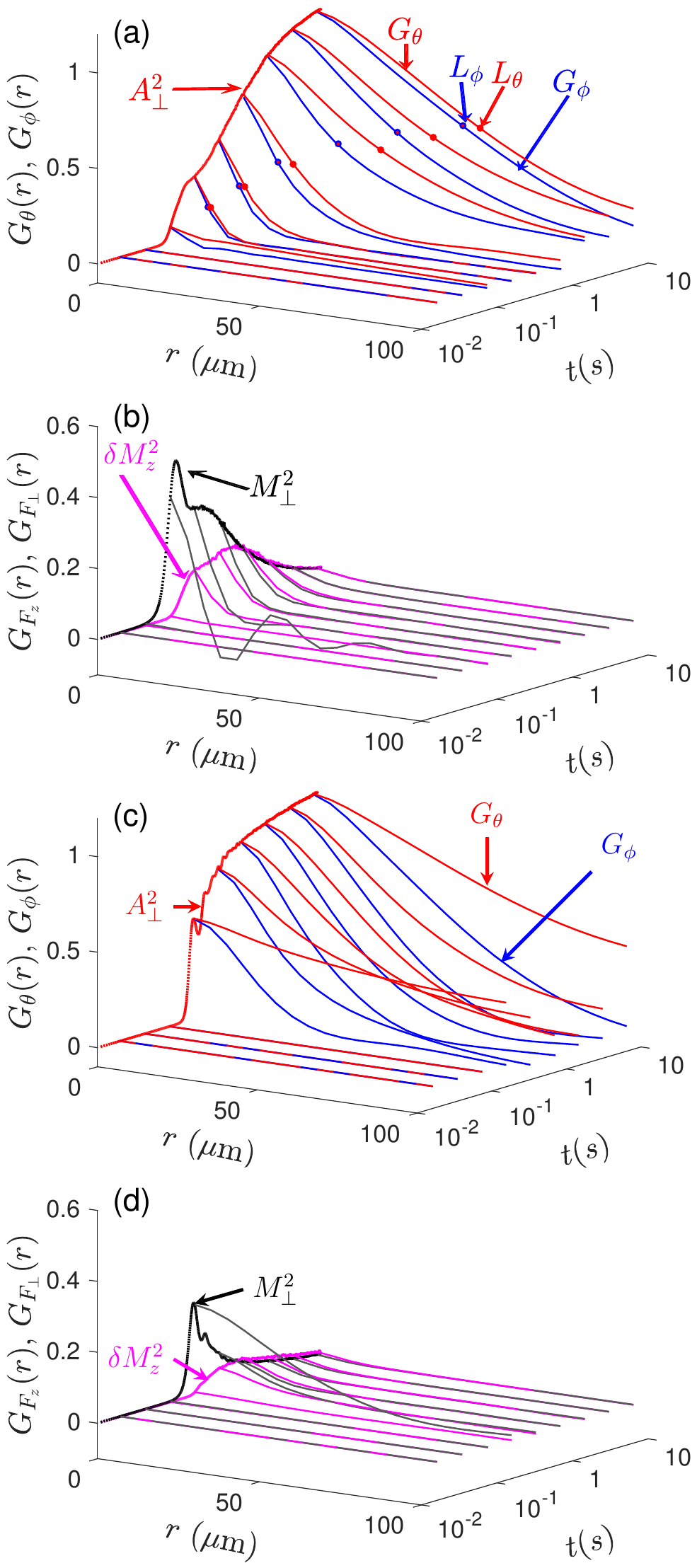} 
\caption{Spatial growth of order following quench. (a) The development of the $G_\theta(r)$ (red line, right of each pair) and $G_\phi(r)$ (blue line, left of each pair) order parameter correlation functions as a function of time for a quench to $q/h=-20\,$Hz. The line at $r=0$ emphasizes the local correlation function evolution and corresponds to $A_\perp^2$ [cf.~Fig.~\ref{fig:localspin}(c)]. The dots indicate the points on the correlation function where it has decayed to 0.5 of its central value, defining the characteristic length scales $L_\theta$ and $L_\phi$.
(b) The development of the $G_{F_z}(r)$ (magenta line) and $G_{F_\perp}(r)$ (grey line) spin correlation functions as a function of time for a quench to $q/h=-20\,$Hz. 
The labeled     lines at $r=0$ emphasize the local correlation function evolution and correspond to $\delta M_z^2$ and $M_\perp^2$ [cf.~Fig.~\ref{fig:localspin}(a) and (b)].  Subplots (c) and (d) are like subplots (a) and (b), respectively, but for a quench to $q/h=-2.7\,$Hz. 
}
\label{fig:corrlfn}
\end{figure}

The results for the spatial evolution of the correlation functions are shown in Fig.~\ref{fig:corrlfn}. Subplots (a) and (c) reveal that the spatial extent of the order increases as time progresses. In contrast the spin-density correlation functions [subplots (b) and (d)]  exhibit transient dynamics following the quench: developing spatial structure inherited from the unstable Bogoliubov modes on the time scale of $t_1\sim100\,$ms, and subsequently decaying.

For shallow quenches, $-8\,\mathrm{Hz} < q/h < 0$, the most unstable modes have wavevectors $k\ll 1/\xi_s$  [e.g.~see Fig.~\ref{BogFig}(a)] and the spin-density correlations can develop on length scales comparable to the size of the system [see $G_{F_\perp}(r)$ in Fig.~\ref{fig:corrlfn}(d)]. In experiments, spatial patterns in the spin density with a scale comparable to the system were observed for shallow quenches (e.g~see Fig.~2 of Ref.~\cite{Kang2017a}). 
We find that this spin order can be transferred to the order parameter [see $G_{\theta}$ in Fig.~\ref{fig:corrlfn}(d)] at early times, although it subsequently decays with the spin correlations before growing again at later times.

\begin{figure}[H] 
\centering
\includegraphics[trim=0 20 0 0,clip=true,width=0.9\linewidth]{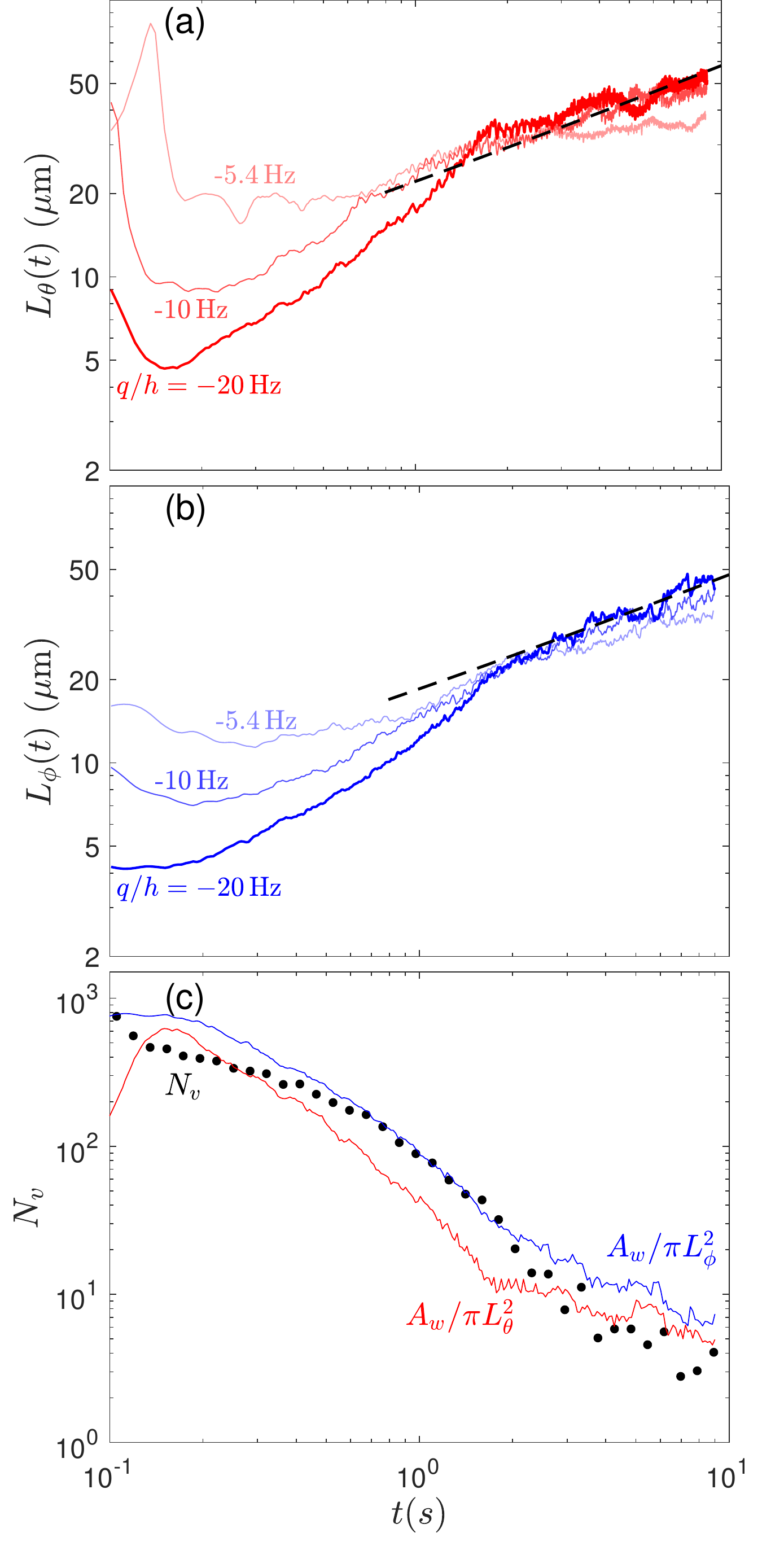} 
\caption{ The evolution of the order parameter length scale (a) $L_{\theta}(t)$  and (b) $L_\phi(t)$ for various $q$ values, as labelled. A fit to the expected growth law $L_g(t)=a[t/\ln(t/t_0)]^{1/2}$ for $t>1\,$s is shown (dashed line) for the $q/h=-20\,$Hz case.  
   (c) The total number of vortices detected in the window region (circles) compared to   $A_w/\pi L_\theta^2$ [red (lower) line] and  $A_w/\pi L_\phi^2$ [blue (upper) line] for the quench to $q/h=-20\,$Hz. 
}
\label{fig:lengthscales}
\end{figure}\vspace{-5mm}

For deeper quenches, $-8\,\textrm{Hz}>q/h$, the most unstable modes have wave vectors $k\sim1/\xi_s$ [e.g.~see Fig.~\ref{BogFig}(b)]  and shorter wavelength spatial patterns emerge in the spin density [see $G_{F_\perp}(r)$ in Fig.~\ref{fig:corrlfn}(b)]. This tends to initialize the order parameter correlation functions with shorter range correlations, and is compatible with the usual initial conditions assumed for coarsening or phase ordering dynamics \cite{Bray1994}. To analyze the ordering dynamics we extract correlation lengths for the order parameter,  defined as the distance over which the correlations decay by half, i.e.~$G_\theta(L_\theta)=0.5G_\theta(0)$ and $G_\phi(L_\phi)=0.5G_\phi(0)$. The length scales $L_\theta$ and $L_\phi$ can be taken as the characteristic size of the spin (nematic) and superfluid domains, respectively.
In Fig.~\ref{fig:lengthscales}(a) and (b) we show the evolution of these length scales for the three deepest quenches. For $q/h=-10\,$Hz and $-20\,$Hz a considerable range of growth is observed. At times $t\gtrsim1\,$s these length scales are $\gtrsim20\,\mu\textrm{m}$ (i.e.~$\sim5\xi_s$), and we may expect that the growth in correlations will become universal. From previous work simulating large uniform 2D systems we have found a universal growth law of $L\sim[t/\ln(t/t_0)]^{1/2}$, and in Fig.~\ref{fig:lengthscales}(a) and (b) we verify that this provides a reasonable fit at late times. Thus we confirm that the experimental system can access the universal regime.

The topological defects of the order parameters are HQVs. We show the evolution of the number of HQVs\footnote{These are detected as phase windings occurring in the $\psi_1$ and $\psi_{-1}$ fields.}
 within   the window region in Fig.~\ref{fig:lengthscales}(c).   
The order parameter correlation lengths scale as the mean distance between the topological defects, which we verify in Fig.~\ref{fig:lengthscales}(c) by showing $1/L_\theta^2$ and $1/L_\phi^2$ for reference.

\section{Conclusion and outlook}\label{Sec:Conclusion}
In this paper we have simulated a recent experiment that  measured the  quench dynamics of a spin-1 antiferromagnetic condensate. We find good qualitative agreement in the observables we compare.  However the initial condensate decay rate (and the subsequent formation of axial magnetization) is slower in our simulations when we use the widely accepted value of $a_s=0.823\,a_0$.  Also, we find that the decay rate saturates for large $|q|$-quenches, whereas the experiments observe the rate to increase with increasing $|q|$.  These differences are largely removed if we use a larger spin-dependent interaction.

Several aspects of the experiment are not accounted for in our theory. For example, heating from the microwave dressing used to control $q$, losses  and heating due to three-body recombination, residual gradients in the magnetic fields, and evaporation of atoms from the finite-depth optical trap. In principle these effects could be added to our classical field formalism, e.g.~3-body recombination would induce new loss and noise terms \cite{Norrie2006a}, and effects of a thermal reservoir could be included via the stochastic GPE formalism \cite{Bradley2014a}. These effects may contribute to the short time dynamics (e.g.~behavior of $\eta$), but clearly have an effect on the late time dynamics (e.g.~occupation of spin sublevels) and it would be interesting to assess their influence on the late-time phase ordering dynamics.  
 
Our study has allowed us to assess the feasibility of studying phase ordering dynamics in the regime of a current spinor experiment. The experiment of Kang \textit{et al.}, is interesting because it is a large system  and is in a quasi-2D regime, whereas other experimental work on antiferromagnetic  phase ordering has been in elongated quasi-one-dimensional traps.  A feature of 2D systems with easy-plane order is that vortices play a fundamental role in the phase ordering and thus exhibit qualitatively different dynamics to one-dimensional systems. While the relevant order parameters were not observed directly in experiments, our results show that monitoring the number of vortices (which are readily observed) can also be used to quantify the late-time phase ordering.

\begin{acknowledgments}
S.~Kang and Y-I.~Shin are acknowledged for useful discussions about their experiment, for providing access to experimental data and bringing Ref.~\cite{Knoop2011a} to our attention.
We acknowledge the contribution of NZ eScience Infrastructure (NeSI) high-performance computing facilities, and support from the Marsden Fund of the Royal Society of New Zealand. 
\end{acknowledgments}


%

\end{document}